\documentclass[twocolumn,showpacs,amsmath,amssymb,aps,prl,floatfix,nofootinbib,superscriptaddress]{revtex4}
\usepackage{graphicx}
\usepackage{dcolumn}
\usepackage{bm}

\begin{document}

\title{Path derivation for a wave scattered model to estimate height correlation function of rough surfaces}
\author{M. Zamani $^{1}$, S. M. Fazeli $^2$, M. Salami $^3$, S. Vasheghani Farahani $^4$, G. R. Jafari $^{1}$
\thanks{Email: g\_jafari@sbu.ac.ir} $^\dag$ \\
{\small $^1$ Department of Physics, Shahid Beheshti University,
G.C., Evin, Tehran 19839, Iran} \\
{\small $^2$ Department of Physics, University of Qom, Qom, Iran} \\
{\small $^3$ Department of Physics, Shahroud Branch, Islamic Azad University, Shahroud, Iran} \\
{\small $^4$ Centre for Plasma Astrophysics, Department of
Mathematics, Katholieke Universiteit Leuven, Celestijnenlaan 200B
bus 2400, B-3001 Heverlee, Belgium } }
\date{\today}

\begin{abstract}

The long standing problem on finding the height correlation function
is studied by the inverse scattering problem. We propose a new
method in the frame work of Kirchhoff theory which we call "path
derivation of scattered wave (PDSW)" in order to obtain an
expression for direct measurements of the height correlation
function. This would provide adequate insight to rough surfaces. The
efficiency of this method is due to the fact that the height
correlation function could be estimated directly by measurements of
the scattered intensity on a suggested path. The model is tested
numerically and an experimental setup is suggested.


\end{abstract}
\maketitle

The importance of rough surfaces in various fields of science and
engineering has been highlighted by many researchers. The area of
study spans from Biology \cite{biol}, Polymer \cite{poly} and  Radio
\cite{rada} Science, Material engineering \cite{Mate}, and Physics
(see \cite{Topi} and references therein), etc. This motivates
researchers specially in Physics to study the various aspects rough
surfaces. The stick slip dynamics of rough surfaces was studied
experimentally \cite{expe} for two surfaces where a hydrodynamic
model was implemented to theoretically study the effects of surface
roughness on viscous incompressible liquids \cite{hydr}. Palasantzas
studied the effect of roughness on ballistic thermal conductance of
a nanosized beam \cite{pala} in addition to effects of the roughness
exponent on friction while a viscous rubber slips in to a rough
surface \cite{pale}. In a detailed study Volokitin and Persson
\cite{Volo} stated that if surface modes for instants adsorbate
vibration modes exist on a surface, thermal radiation would be
coherent and due to the electromagnetic oscillations on the surface,
radiative heat transfer would be experienced.

The topography of any surface could be obtained by mechanical
devices such as scanning probe microscopy, SPM \cite{STM, AFM} e.g.
in application to semiconducting carbon nanotube transistors
\cite{AFMA}. But despite the wide range of applicability and
benefits of SPM, the tip convolution may still remain as the main
challenge for SPM \cite{Newj}, in addition to the limitations that
SPM has in changing the scale of scanning. Optical techniques would
overcome this limitation without scratching the surface. Optical
techniques could also manage to determine the scale of observation
with the incident light wave-length. The mostly implemented method
in the optical technique is the scattering method where the
roughness would play a major role on the scattering of waves
\cite{Ogilvy}. Interesting work in the context of wave scattering
has been carried out in application to Brewster's scattering angle
\cite{Brew} and the Rayleigh hypothesis \cite{Reig}. Since the
wave-length could be either smaller or greater than the height
fluctuations, the two scale theory was proposed \cite{Rjafa}. In a
further study, effects of interference of two beam scattering was
considered \cite{Jafa}. All these plus the fact that studying the
scattered wave would provide very useful information about the
surface itself \cite{Back}, would motivate studying the inverse
scattered problem.

Inverse scattering techniques were implemented
\cite{pierri,liseno,qing1,qing2,ferraye} to measure the statistical
properties of rough surfaces. Although after quite a long time since
optical techniques were born, and many parameters like; root mean
square of height \cite{Dast}, probability density function (PDF)
\cite{RJaf}, correlation length \cite{Helm} etc. where estimated,
not much progress has been achieved in order to find the most
important parameter  of the surface which is the height correlation
function. This issue is the main aim of this work. The framework
implemented is wave scattering by Kirchhoff theory approach
\cite{beck,kong,fung,Ogilvy}. The Kirchhoff theory is an
electromagnetic theory which treats any point on a scattering
surface as a part of an infinite plane, parallel to the local
surface tangent \cite{Ogilvy}. In this work a model is proposed to
estimate the height correlation function of a random surface which
we name; "path derivation of scattered wave". Taking into account
this model enables experimentalists to obtain the height correlation
function simply by only measuring the intensity of the scattered
light in a special path. This technique would dramatically simplify
the estimation of the height correlation function due to the fact
that only one parameter (the scattered intensity) needs to be
measured.

In Kirchhoff theory, the field of incident monochromatic wave may be
written as $\psi^{inc}(r)=\mathrm{exp}(-i k_{inc}.r)$, where $k$ and $r$
indicate the wave number and position respectively. In Fig. 1 a beam
is shown incident to a rough surface with dimensions
$-X\leq x_{0}\leq X$, $-Y\leq y_{0}\leq Y$ with a reflection
coefficient $R_0$ equal to $-1$ corresponding to the Dirichlet
boundary condition.
\begin{figure}[t]
\includegraphics[width=8cm,height=6cm,angle=0]{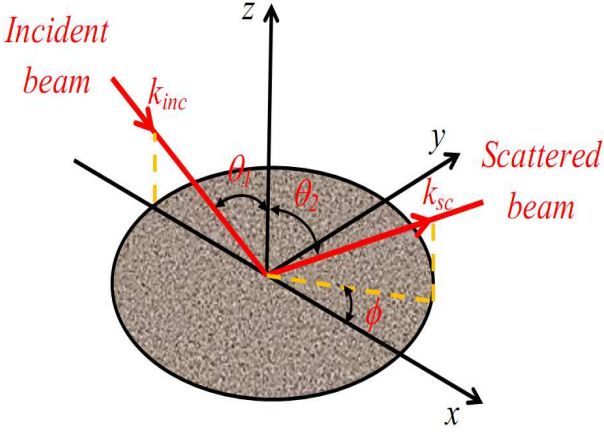}
\caption{The geometry used for wave scattering from a rough surface.} \label{Fig1}
\end{figure}
 The total scattered field over the mean
reference plane $A_{M}$, is given by \cite{Ogilvy}
\begin{eqnarray}
 \psi^{sc}(r) & &= \frac{ike^{ikr}}{4\pi r}\int_{S_{M}}
\left(a\frac{\partial h}{\partial x_{0}}+b\frac{\partial h}{\partial
y_{0}}-c\right) \nonumber\\
& & \times\mathrm{exp}\left(ik[Ax_{0}+By_{0}+
Ch(x_{0},y_{0})]\right)dx_{0} dy_{0},
\end{eqnarray}
where
\begin{eqnarray}
A&=&\sin \theta_{1}-\sin \theta_{2} \cos \phi, \nonumber \\
B&=&-\sin \theta_{2} \sin \phi, \nonumber \\
C&=&-(\cos \theta_{1} +\cos \theta_{2}), \nonumber \\
a&=&\sin \theta_{1}(1-R_{0})+\sin \theta_{2}\cos\phi(1+R_{0}), \nonumber \\
b&=&\sin \theta_{2} \sin \phi(1+R_{0}), \nonumber \\
c&=&\cos \theta_{2}(1+R_{0})-\cos \theta_{1}(1-R_{0}).
\end{eqnarray}
The total scattered intensity for isotropic surfaces could be
considered as sum of the diffused and coherent parts, the diffused
part is \cite{Ogilvy}
\begin{eqnarray}
\label{Id}
 <I_{d}> = \frac{k^{2}F^{2}}{2\pi r^{2}}A_{M}\mathrm{exp}(-g) \int
_{0}^{\infty}J_{0}\left(kR\sqrt{A^{2}+B^{2}}\right)\nonumber\\
\times\left[\mathrm{exp}(gCor(R))-1\right] R dR,
\end{eqnarray}
where $Cor(R)=\frac{\langle h(\textbf{x}+\textbf{R})
h(\textbf{x})\rangle}{\sigma^2}$ is the height-height correlation
function and $\langle h(\textbf{x}) \rangle=0$,
$g=k^2\sigma^2(\mathrm{cos}\theta_1+\mathrm{cos}\theta_2)^2$ is the
roughness criteria, and $\sigma=\sqrt{\langle (h(\textbf{x}) -
\langle h(\textbf{x}) \rangle )^2 \rangle }$ is height root means
squared of the rough surface. It is more convenient to write
equation (\ref {Id}) in the form
\begin{eqnarray}
\label{Idnew}
 \langle \widetilde{I_{d}} \rangle &=& \int
_{0}^{\infty}J_{0}\left(kR\sqrt{A^{2}+B^{2}}\right)\nonumber\\
 & & \times [\mathrm{exp}(gCor(R))-1] R dR,
\end{eqnarray}
with $I_{d}/\widetilde{I_{d}}=\frac{k^{2}F^{2}}{2\pi r^{2}}A_{M}
\mathrm{exp}(-g)$, where from hereafter the overtilde is omitted.
Equation (\ref{Idnew}) shows the diffused intensity in terms of the
height correlation function, where the idea is to find the height
correlation function in terms of the diffused intensity. This would
be obtained by solving equation (\ref{Idnew}) making use of the
saddle point approximation. Attention must be focused on a few
issues; we are considering rough surfaces were $g\gg1$, this means
that the conditions, $k\sigma \gg1$ and small $\theta_1$ and
$\theta_2$ must be fulfilled.

- In highly rough surfaces, $g\gg1$, the contribution of the
coherent term could be neglected in comparison to the contribution
of the diffused term, this means that the diffuse intensity $I_{d}$
would be equal to the total intensity $I$ \cite{Ogilvy}, expressed
as
\begin{eqnarray}
\label{Itotal}
 {I}= \int
_{0}^{\infty}J_{0}\left(kR\sqrt{A^{2}+B^{2}}\right)
\exp(gCor(R))RdR.
\end{eqnarray}
- In order to satisfy the conditions of the saddle point
approximation, the integrand must tend to zero on the integral
limits. This condition is not satisfied by Eq. (\ref{Itotal}),
persuading us to differente both side of Eq. (\ref{Itotal}) in
respect to $g$ resulting in
\begin{eqnarray}
\label{Idnewcoefficient}
 \frac {d}{dg} I &=& \int
_{0}^{\infty}Cor(R)J_{0}\left(kR\sqrt{A^{2}+B^{2}}\right)\nonumber\\
& & \times\exp(gCor(R))RdR.
\end{eqnarray}
Note that the derivative has been taken on a path where the Bessel
function $J_{0}(kR\sqrt{A^{2}+B^{2}})$ stays constant in respect to
the variable $g$. It could readily be seen that the term
$J_{0}(kR\sqrt{A^{2}+B^{2}})$ depends on the wave-length, incident
and scattered angles. The Bessel function is kept constant with
respect to the variations of $g$. This is done by keeping the
wave-length and incident angle constant and only letting the
scattered angles vary in a way to keep the $A^2+B^2$ constant, see
Fig. 2. In Fig. 2, three typical paths corresponding to three
different constant values of $A^{2}+B^{2}$ is shown.

Making use of the definition of the logarithmic function we obtain
\begin{eqnarray}
\label{Iderivative}
 \frac {d}{dg} I & &= \int
_{0}^{\infty}J_{0}\left(kR\sqrt{A^{2}+B^{2}}\right) \nonumber\\
\times & & \exp\left[g\left(Cor(R)+\frac{
\ln(R)}{g}+\frac{\mathrm{ln}(Cor(R))}{g}\right)\right] dR.
\end{eqnarray}
Equation (\ref{Iderivative}) would satisfy the conditions for the
saddle point approximation, where the argument of the Bessel
function must stay constant in respect to $g$. According to Fig. 2,
the variation of $\theta_{2}$ and $\phi$ would provide the path
where the intensity is measured on. Note that in plotting the curves
(paths) of Fig. 2, the Bessel function stays constant with respect
to the variations of $g$.

- To solve the integral of Eq. (\ref{Iderivative}) by the saddle
point approximation, Eq. (\ref{Iderivative}) needs to be written in
a form so how that the extremum would be inside the integral limits
and not on the integral limits. By implementing a new variable
$u=\ln(R)$, the lower limit of the integral could be changed to
$-\infty$. Hence, Eq.(\ref{Iderivative}) would be
\begin{eqnarray}
\label{Iderivativeu}
 \frac {d}{dg} I &=& \int
_{-\infty}^{\infty}J_{0}\left(k \mathrm{exp(u)}\sqrt{A^{2}+B^{2}}\right) \nonumber\\
& &
\times\exp\left[g\left(\widetilde{Cor}(u)+\frac{2u}{g}+\frac{\ln(\widetilde{Cor}(u))}{g}\right)\right]du.
\end{eqnarray}
Since the intensity and intensity variations are needed on the path,
and depending on the chosen constant $A^{2}+B^{2}$, one could refine
the correlation. This could be obtained due to the fact that all
paths would give a unique height correlation function. This would
suggest a new technique for experimentalists to test their obtained
results by choosing specific incident angles, incident wave-lengths,
to keep the desired $J_0$ constant in respect to $g$. In other words
to keep $J_{0}$ constant, the parameter $A^2+B^2$ must stay
constant. In this way a new path may be suggested for finding the
intensity experimentally. Fig. 2 shows three paths obtained for
three constant values of $A^2+B^2$, $\theta_1=20^0$ and
$\lambda=500\, nm$.

\begin{figure}[t]
\includegraphics[width=8cm,height=5.5cm,angle=0]{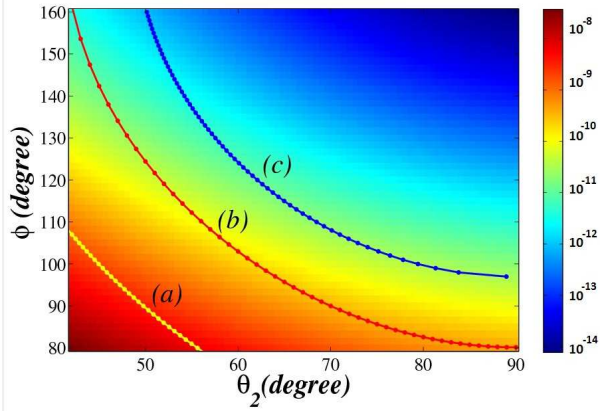}
\caption{ (Color online) The curves are typical paths on which the
Bessel function or $A^2+B^2$ stays constant with fixed $k$ as $g$
varies. The plane of consideration is the diffused intensity based
on the scattered angles $\theta_2$ and $\phi$ for a Gaussian
correlation function. The values for the parameters are;
$\theta_{1}=20^{\circ}$, $\sigma=50 nm$ and $\lambda=500 nm$. The
paths correspond to (a) $A^{2}+B^{2}=0.7$ (b) $A^{2}+B^{2}=1$ (c)
$A^{2}+B^{2}=1.2$.} \label{fig2}
\end{figure}
In this stage Eq. (\ref{Iderivativeu}) complies the saddle point
approximation by
\begin{eqnarray}
\label{FU}
f(u) &=& \widetilde{Cor}(u)+\frac{2u}{g}+\frac{\ln(\widetilde{Cor}(u))}{g},\nonumber\\
f^{\prime}(u^*) &=& \widetilde{Cor}^{\prime}(u^*)+\frac{2}{g}+\frac{\widetilde{Cor}^{\prime}(u^*)}{g\widetilde{Cor}(u^*)}=0,\nonumber\\
f^{\prime\prime}(u^*) &=&
\widetilde{Cor}^{\prime\prime}(u^{*})+\frac{1}{g}\frac{\widetilde{Cor}^{\prime\prime}(u^{*})\widetilde{Cor}(u^{*})-\widetilde{Cor}^{\prime2}(u^*)}{\widetilde{Cor}^{2}(u^{*})},
\end{eqnarray}
where, $u^*$ is the desired saddle point. Solving the integral in
equation (\ref{Iderivativeu}) by the saddle point approximation
gives
\begin{eqnarray}
\label{IComplicated}
 \frac{d}{dg}I &=& \sqrt{2\pi}J_{0}\left(k \mathrm{exp}(u^{*})\sqrt{A^{2}+B^{2}}\right)\nonumber\\
 & & (gf^{\prime\prime}(u^*))^{-\frac{1}{2}}
 \mathrm{exp}(g f(u^{*})),
\end{eqnarray}
due to the complexity of the RHS of Eq. (\ref{IComplicated}), an
explicit version for the height correlation function may not be
obtained without the use of the second derivative of Eq.
(\ref{Iderivativeu})
\begin{eqnarray}
\label{IDerivative}
\frac{d^2}{dg^{2}}I &=& \widetilde{Cor}(u^{*})\sqrt{2\pi}J_{0}\left(k \mathrm{exp}(u^{*})\sqrt{A^{2}+B^{2}}\right)\nonumber\\
& & (gf^{\prime\prime}(u^*))^{-\frac{1}{2}} \mathrm{exp}(gf(u^{*})),
\end{eqnarray}

Equation (\ref{IDerivative}) is an expression in terms of
$\widetilde{Cor}(u)$ in the extremum point which itself is dependent
on $g$. This means that changing $g$ by differing the wave-length,
standard deviation, incident and scattering angles, the correlation
function would vary. This enables us to obtain the functionality of
the correlation on $g$ instead of just having a specific point. By
dividing both sides of Eqs. (\ref{IComplicated}) and
(\ref{IDerivative}), the expression for $\widetilde{Cor}(u^*)$ is
obtained
\begin{equation}
\label{uexplicit}
\widetilde{Cor}(u^*)=\frac{\frac{d^2}{dg^2}[I]}{\frac{d}{dg}[I]}=W(g).
\end{equation}
Equation (\ref{uexplicit}) shows the relation between
$\widetilde{Cor}(u^*)$ and $g$. Due to the fact that $W(g)$ could be
observed experimentally, Eq. (\ref{uexplicit}) would give the
correlation function. But the issue which still remains is the value
of $u^*$.

From the second expression of Eq. (\ref{FU}) an explicit relation
between the derivative of the height correlation and $g$ for the
limiting case of highly rough surfaces ($g\gg1$) is obtained
\begin{equation}
\label{neglect}
 \widetilde{Cor}^{\prime}(u^*)=-\frac{2}{g+\frac{1}{\widetilde{Cor}(u*)}}
\simeq -\frac{2}{g}.
\end{equation}

The derivative of Eq. (\ref{uexplicit}) with respect to $g$ would
show the functionality of $u^*$ in terms of $g$
\begin{equation}
\frac{d\widetilde{Cor}(u^*)}{dg}= \frac{du^*}{dg}
\widetilde{Cor}^{\prime}(u^*)=\frac{dW(g)}{dg},
\end{equation}
and by using Eq. (13)
\begin{equation}
u^*(g) = - \int \frac{g}{2} \frac{dW(g)}{dg} dg,
\end{equation}
which with partial integration, it could be obtained $u^{*}(g)$
\begin{equation}
\label{ustarg}
u^{*}(g)=-\frac{gW(g)}{2}+\int \frac{W(g)}{2}dg.
\end{equation}
Equation (\ref{ustarg}) gives an expression for $u^*$ on $g$. The
relation between $\widetilde{Cor}(u^{*})$ with $g$ is already shown
in Eq. (\ref{uexplicit}), so the dependence of the height
correlation function on $u$ may readily be obtained. Since
$u=\ln(R)$, the dependence of the height correlation function on $R$
may be obtained.

The procedure could be illustrated as follows; incident light with a
specific angle to a rough surface with a constant wave length
(preferably small) scatters with a verity of angles. The measured
intensity for each scattered angle on the desired path needs to be
differentiated in respect to $g$ (which depends on the scattered
angle) twice. Then the ratio of the second derivative and the first
derivative is obtained in order to find $W(g)$. Substituting $W(g)$
in Eq. (\ref{ustarg}) the parameter $u^{*}(g)$ is obtained, keeping
in mind that $W(g)$ is the height correlation of $u^{*}$. Having the
dependence or in other words variations of $\widetilde{Cor}(u^{*})$
and $u^{*}$ on $g$ the height correlation on $R$ is obtained.

In Fig. 3 the applicability and robustness of this technique is
illustrated. Where as an example we estimate the known exponential
correlation function with the PDSW model and compare it with the
original cases showing a very good consistency. The parameters for
this estimation are; $A^2+B^2=0.8$, $\theta=20^{\circ}$,
$\sigma=0.5$ and $k \sigma=1.2$.

\begin{figure}[t]
\includegraphics[width=9cm,height=6cm,angle=0]{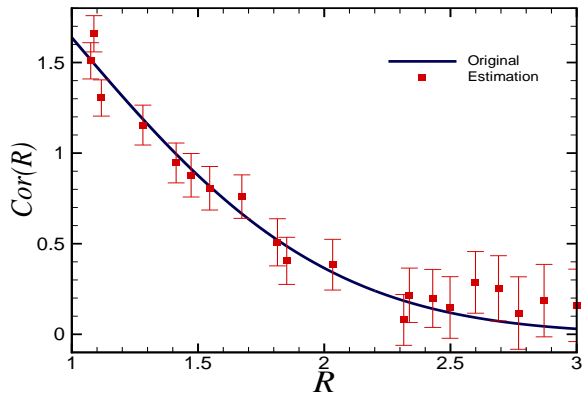}
\caption{Comparison of the correlation function estimated by the
PDSW model and the original exponential correlations. The parameters
for this estimation are; $A^2+B^2=0.8$, $\theta_1=20^{\circ}$,
$\sigma=0.5$ and $k \sigma=1.2$.} \label{Fig3}
\end{figure}

The applicability of the method introduced in this work for
experimentalists although has a very simple procedure, would give
the height correlation function of the surface. The height
correlation function is the most important feature in order to
understand the physics of a rough surface which is applicable to
various fields of physics. "The path derivative scattered wave"
introduced in this work would prove as an easy to use technique for
experimentalists since only by measuring the intensity of the
scattered light on a specific path, the height correlation function
of a specific rough surface could be obtained.


\end{document}